\documentclass[12pt,letterpaper,copyedit]{article}
\usepackage{osajnl2}
\usepackage{amsmath}

\begin{document}


\title{Photonic realization of the relativistic Dirac oscillator}


\author{S. Longhi}
\address{Dipartimento di Fisica,
Politecnico di Milano, Piazza L. da Vinci 32, I-20133 Milano, Italy}
\begin{abstract}
A photonic realization of the Dirac oscillator (DO), i.e. of the
relativistic extension of the quantum harmonic oscillator, is
proposed for light propagation in fiber Bragg gratings. Transmission
spectra clearly show the existence of electron and positron bound
states of the DO, corresponding to resonance modes above and below
the Bragg frequency, as well as the asymmetry of the energy spectrum
for electron and positron branches.
\end{abstract} \ocis{000.2658,060.3735}



\noindent
 The relativistic extension
of the quantum harmonic oscillator, the so-called Dirac oscillator
(DO) \cite{D1,D2,D3}, provides a paradigmatic and exactly solvable
model of relativistic quantum mechanics. Originally proposed in
quantum chromodynamics in connection to quark confinement models in
mesons and baryons \cite{D4}, the DO has received great interest in
relativistic many-body theories and supersymmetric relativistic
quantum mechanics (see \cite{D2,D3,D5,D6,D7} and references
therein). The DO model is obtained from the free Dirac equation by
the introduction of the external potential via a non-minimal
coupling \cite{D1,D3,D7}. Since the resulting equation is linear in
both momentum and position operators,  in the nonrelativistic limit
a Schr\"{o}dinger equation with a quadratic potential is then
obtained. In spite of the great amount of theoretical studies, the
DO model in relativistic quantum mechanics and particle physics
remains far from any experimental consideration. Recently, an
increasing interest has been devoted to the investigation of quantum
or classical systems capable of providing an accessible laboratory
tool to test some peculiar phenomena rooted in the Dirac equation,
such as Klein tunneling and Zitterbewegung \cite{Greiner}. Quantum
and classical analogs of such two phenomena have been investigated
in graphene, trapped ions, photonic crystals and metamaterials (see,
e.g., \cite{AN1,AN2,AN3,OC1,OC2,OC3,OC4} and references therein),
with the first experimental observations of Klein tunneling and
Zitterbewegung reported in Refs.\cite{AN1,AN2,AN3}. Such experiments
motivate the search for experimental systems that could simulate a
relativistic DO. Recently, interesting insightful connections
between the relativistic DO and the Jaynes-Cummings model of quantum
optics have been pointed \cite{JC1,JC2,JC3}, and an experimental
proposal of a two-dimensional DO using a single trapped ion has been
presented \cite{JC2}. \\
It is the aim of this Letter to propose a classic wave optics
analogue of the one-dimensional relativistic Dirac oscillator based
on light propagation in engineered fiber Bragg gratings (FBGs). Let
us consider light propagation in a FBG \cite{Erdogan} with a
longitudinal refractive index $ n(z)=n_0+\Delta n \; h(z) \cos[2 \pi
z / \Lambda + \phi(z)]$, where $n_0$ is the effective mode index in
absence of the grating, $\Delta n \ll n_0$ is a reference value of
the index change of the grating, $\Lambda$ is the nominal grating
period defining the Bragg frequency $\omega_B=\pi c/(\Lambda n_0)$,
$c$ is the speed of light in vacuum, and $h(z)$, $\phi(z)$ describe
the amplitude and phase profiles, respectively, of the grating. To
study Bragg scattering of counterpropagating waves at frequencies
close to $\omega_B$, let $E(z,t)=\{ u(z,t) \exp(-i \omega_B t +2 \pi
i z n_0 / \lambda_0)+ v(z,t) \exp(-i \omega_B t -2 \pi i z n_0 /
\lambda_0)+c.c. \}$ be the electric field in the fiber, where
$\lambda_0=2 n_0 \Lambda$ is the Bragg wavelength. The envelopes $u$
and $v$ of counterpropagating waves [see Fig.1(a)] satisfy the
coupled-mode equations \cite{Erdogan,Poladian}
\begin{eqnarray}
 \left[ \partial_z + (1/v_g) \partial_t
\right] u & = & iq(z) v \\
 \left[ \partial_z -(1/v_g) \partial_t  \right] v & = & -iq^*(z)u
\end{eqnarray}
where $q(z)=( \pi \Delta n / \lambda_0) h(z) \exp[i \phi(z)]$ is the
complex scattering potential and $v_g \sim c/n_0$ is the group
velocity at the Bragg frequency. To highlight the analogy between
light propagation in a FBG with suitable amplitude and and phase
profiles, and the one-dimensional relativistic DO, let us introduce
the dimensionless variables $x=z/Z$ and $\tau=t/T$ , with
characteristic spatial and time scales $Z=\lambda_0/(\pi \Delta n)$
and $T=Z/v_g$, and the new envelopes $\psi_{1,2}(z)=[u(z) \mp v(z)]/
\sqrt 2$. In this way, Eqs.(1-2) can be cast in the Dirac form
\begin{equation}
i \partial_{\tau} \psi=\sigma_x \left\{ p_x -if(x) \sigma_z \right\}
\psi + \sigma_z m(x) \psi
\end{equation}
for the spinor wave function $\psi=(\psi_1,\psi_2)^T$, where
$\sigma_{x}$ and $\sigma_{z}$ are the Pauli matrices, defined by
\begin{equation}
\sigma_x= \left(
\begin{array}{cc}
0 & 1 \\
1 & 0
\end{array}
\right) \; , \; \; \sigma_z= \left(
\begin{array}{cc}
1 & 0 \\
0 & -1
\end{array}
\right),
\end{equation}
$p_x=-i (d/dx)$, and where we have set
\begin{equation}
m(x)=h(x) \cos[\phi(x)] \; , \; \; f(x)=-h(x) \sin[\phi(x)].
\end{equation}
In its present form, Eq.(3) is analogous to the one-dimensional
Dirac equation \cite{Greiner}, written in atomic units
($\hbar=c=1$), with a space dependence mass $m$ and with the
momentum operator $p_x$ substituted with $p_x-if(x) \sigma_z$. The
space dependence of the particle mass $m$ is known to describe the
particle interaction with a scalar Lorentz potential, whereas the
substitution $p_x \rightarrow p_x-if(x) \sigma_z$ corresponds to a
non-minimal coupling which is essential to describe the relativistic
DO (see, for instance, \cite{D7}). To realize the one-dimensional
analog of the DO, let us choose the amplitude $h$ and phase $\phi$
profiles of the grating such that $h \cos \phi=m_0$ and $h \sin
\phi=-f(x)= \omega_s m_0 x$, i.e. (see Fig.1)
\begin{equation}
h(x)=m_0 \sqrt{1+(\omega_s x)^2} \; , \; \; \phi(x)= {\rm
atan}(\omega_s x),
\end{equation}
where $m_0$ and $\omega_s$ are two arbitrary constants,
corresponding to the particle rest mass and oscillation frequency of
the DO in the non-relativistic limit \cite{D7}. Analytical
expressions of the energy spectrum for the one-dimensional DO and of
corresponding bound states can be derived following a standard
procedure detailed e.g. in Ref.\cite{D7}. Let us search for a
solution to Eq.(3) of the form $\psi(x,\tau)=(\psi_+(x),\psi_-(x))^T
\exp(-i \delta \tau)$, where $\delta$ is the energy eigenvalue of
the Dirac Hamiltonian and $m(x)$, $f(x)$ are defined according to
Eqs.(5) and (6). The functions $\psi_{+}$ and $\psi_{-}$ are then
found to satisfy the harmonic oscillator equation
\begin{equation}
-\frac{1}{2m_0}\frac{d^2 \psi_{\pm}}{dx^2}+\frac{1}{2}m_0 \omega_s^2
x^2 {\psi_{\pm}}=\frac{\delta^2-m_0^2 \mp m_0 \omega_s}{2m_0}
\psi_{\pm}.
\end{equation}
The eigenvalues of Eq.(7) are those of the non-relativistic harmonic
oscillator. Assuming for the sake of definiteness $\omega_s>0$, the
positive-energy spectrum (electron branch) of the DO is thus given
by
\begin{equation}
\delta_n= \sqrt{m_0^2+2m_0 \omega_s(1+n)} \; , \; \; n=0,1,2,3,...
\end{equation}
whereas the negative-energy spectrum (positron branch) is given by
\begin{equation}
\delta_n= -\sqrt{m_0^2+2m_0 \omega_s n} \; , \; \; n=0,1,2,3,...
\end{equation}
The corresponding eigenfunctions $\psi_{\pm}(x)$ can be simply
expressed in terms of Hermite polynomials multiplied by a Gaussian
function. It is worth pointing out two important properties of the
spectrum of the one-dimensional DO. First, the non-relativistic
limit of the DO is attained by considering the positive-energy
branch of the spectrum in the limit $\omega_s/m_0 \rightarrow 0$; in
this limit, from Eq.(8) one obtains at leading order
$\delta_n=m_0+\omega_s(1/2+n)+(\omega_s/2)$ ($n=0,1,2,3...$). The
first two terms appearing in this expression of the energy have a
simple physical meaning: $m_0$ is the rest energy of the particle
whereas $\omega_s(1/2+n)$ are the usual quantized energy levels
 of the non-relativistic harmonic oscillator (please note
that we used atomic units, i.e. $\hbar=c=1$). The third term
$(\omega_s/2)$ is responsible for an additional energy shift of the
zero-point energy, which is a characteristic feature of the DO (also
found in the three-dimensional DO; see for instance \cite{D3,D5}). A
second property of the DO, which readily follows from an inspection
of Eqs.(8) and (9), is that the negative (positron) energy spectrum
is not obtained from the positive (electron) energy spectrum by sign
reversal ($\delta \rightarrow -\delta)$, the positron branch
possessing an additional bound state with energy $\delta=-m_0$. In
our FBG realization of the DO, bound-states with positive and
negative energies should correspond to trapped light states in the
FBG with resonance frequencies above ($\delta>0$) and below
($\delta<0$) the Bragg frequency $\omega_B$, respectively.  In a FBG
of finite length, the ideal amplitude profile $h(x)$, defined by
Eq.(6), must be truncated, i.e. one has $h(x)=0$ for $|x|>L/(2Z)$,
where $L$ is the grating length (see Fig.1). The effect of grating
truncation is twofold. First, the bound states of the DO become
actually resonance modes with a finite lifetime, which should be
thus observable as narrow transmission peaks embedded in the stop
band of the grating (see Fig.2 to be commented below). Second, the
number of resonance modes sustained by the grating is finite owing
to grating truncation. As an example, Fig.2(a) shows a typical
transmission spectrum [power transmission versus normalized
frequency detuning $\delta=(\omega-\omega_B)T$], as obtained by
standard transfer matrix methods \cite{Erdogan}, of a FBG with
length $L/Z=10$ for parameter values $m_0=1$ and $\omega_s=0.5$.
Figure 2(b) shows the corresponding reflection-band diagram (dashed
area) of the grating in the $(x,\delta)$ plane \cite{Poladian},
together with the electron ($\delta>0$) and positron ($\delta<0$)
levels of the DO, as predicted by Eqs.(8) and (9); a snapshot of
normalized intensity profiles
$|\psi_{+}|^2+|\psi_{-}|^2=|u|^2+|v|^2$ for a few low-order DO bound
states are also depicted. According to Ref.\cite{Poladian}, the band
reflection diagram shows the local stop band of the grating, as a
function of position $x$, which is determined by the inequality $|
\delta-(1/2) (d \phi / dx)| < |h(x)|$. Note that the transmission
peaks in Fig.2(a), embedded in the stop band of the FBG, occur
precisely at the values $\delta_n$ predicted by Eqs.(8) and (9).
Note also the asymmetry of the transmission spectrum around
$\delta=0$, with an additional resonance in the positron
($\delta<0$) branch of the spectrum with no counterpart in the
electron ($\delta>0$) branch. According to the band reflection
diagram of Fig.2(b), the number of bound states of the DO sustained
by the truncated FBG are about 6 in each branch. For the sake of
clearness, only 3 modes in the electronic branch, and 4 modes in the
positron branch, are depicted in the figure. Note that, as expected,
the resonance widths increase for higher-order modes (i.e. for
increasing quantum number $n$) because of the narrowing of the gap
regions that sustain light trapping for such modes. Note also that
additional oscillations in the transmission spectrum, outside the
stop band of the grating, are clearly visible in Fig.2(a). However,
such oscillations do not correspond to resonance states of the DO,
rather they arise because of an impedance mismatch between the
grating ($|z|<L/2$) and non-grating ($|z|>L/2$) regions of the
fiber, like in an ordinary non-apodized FBG (see, for instance,
\cite{Erdogan}). In physical units, for typical parameter values
$n_0=1.45$, $\Delta n=1 \times 10^{-4}$ and $\lambda_0 =1560$ nm
which apply to FBGs used in optical communications, the spatial
($Z$) and temporal ($T$) scales for the example in Fig.2 are
$Z\simeq 5$ mm and $T \simeq 24$ ps, respectively. Hence, in
physical units the grating length is $L \simeq 5$ cm, the unit scale
of (non-angular) frequency detuning in Fig.2 is $1/(2 \pi T) \simeq
6.6 \; {\rm GHz}$, and the maximum index change requested to realize
the FBG, reached at the FBG edges $z= \pm L/2$, is $\Delta n \sqrt{
1+(\omega_s L/2Z)^2} \simeq 2.69 \times 10^{-4}$. Such a nonuniform
FBG should be easily manufactured with currently available FBG
technology based on UV continuous laser writing \cite{fab}.
According to Fig.2(a) and following the previous discussion, in an
experiment the resonant states of the DO could be simply detected
from spectrally-resolved transmission measurements. Like in any FBG
structure supporting trapped modes (e.g. Fabry-Perot or
phase-shifted FBGs), the transmission peaks of high-Q resonant modes
may be strongly reduced by unavoidable intrinsic losses in the FBG
(see, for instance, \cite{loss}). As an example, Fig.2(c) shows the
transmission spectrum of the FBG of Fig.2(a), but assuming a power
loss rate $\alpha=0.5 \; {\rm m}^{-1}$ in the fiber (value taken
from Ref.\cite{loss}). Note that, in spite of the great peak
reduction,
the resonant states as well as the asymmetry of the spectrum are still clearly visible.\\
In conclusion, a photonic realization of a Dirac oscillator, based
on light propagation in FBGs, has been proposed. Electron and
positron bound states of the DO should be clearly visible in
spectrally-resolved transmission measurements as resonances above
and below the Bragg frequency.\\
\\
The author acknowledges support by the Italian MIUR (PRIN 2008
project "Analogie ottico-quanstistiche in strutture fotoniche a
guida d'onda").

 Author E-mail address: longhi@fisi.polimi.it

\newpage

{\bf List of Figure Captions.}\\
\\
{\bf Fig.1.} (a) Schematic of a FBG. (b) and (c) Example of FBG
amplitude and phase profiles that realize the analog of the Dirac
oscillator ($m_0=1$, $\omega_s=0.5$, $L/Z=10$).\\
\\
{\bf Fig.2.} (Color online) (a) Numerically-computed power
transmission (dB units) of a lossless FBG with amplitude and phase
profiles shown in Fig.1, and (b) corresponding reflection band
diagram (dashed area). In (b) a few low-order intensity profiles of
trapped modes in the electron and positron branches are also
depicted. (c) Same as (a), but for a lossy FBG (power loss rate
$\alpha=0.05 \; {\rm m}^{-1}$).

\newpage

\begin{figure}[htb]
\centerline{\includegraphics[width=8.2cm]{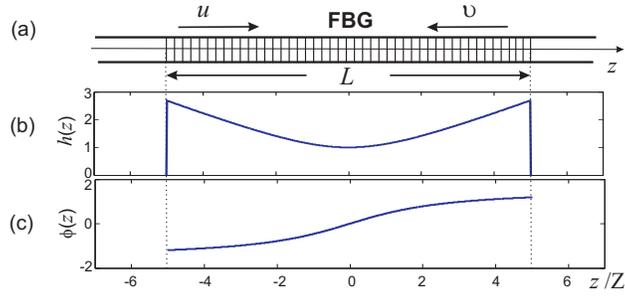}} \caption{(a)
Schematic of a FBG. (b) and (c) Example of FBG amplitude and phase
profiles that realize the analog of the Dirac oscillator ($m_0=1$,
$\omega_s=0.5$, $L/Z=10$).}
\end{figure}

\begin{figure}[htb]
\centerline{\includegraphics[width=8.2cm]{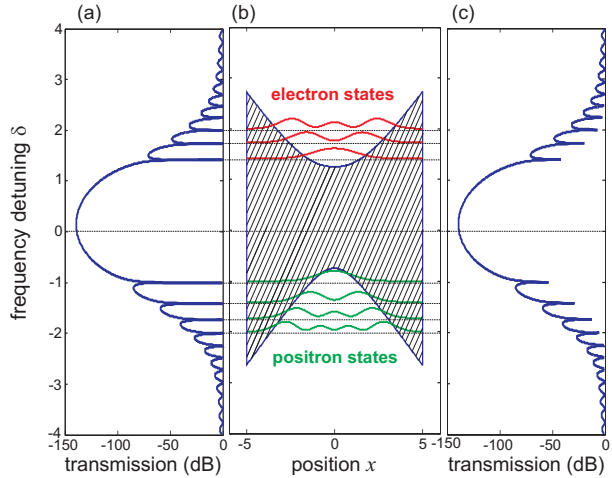}} \caption{
(Color online) (a) Numerically-computed power transmission (dB
units) of a lossless FBG with amplitude and phase profiles shown in
Fig.1, and (b) corresponding reflection band diagram (dashed area).
In (b) a few low-order intensity profiles of trapped modes in the
electron and positron branches are also depicted. (c) Same as (a),
but for a lossy FBG (power loss rate $\alpha=0.05 \; {\rm
m}^{-1}$).}
\end{figure}

\end{document}